\documentclass[journal,draftclsnofoot,onecolumn,12pt,twoside]{IEEEtran}
\normalsize

\ifCLASSINFOpdf
  \usepackage[pdftex]{graphicx}
  \graphicspath{{figures/}}
  \DeclareGraphicsExtensions{.pdf,.jpeg,.png}
\else

\fi

\usepackage{xcolor}
\usepackage{amsmath,amssymb}
\usepackage{physics}
\usepackage[utf8]{inputenc}
\usepackage{subfig} 
\usepackage{soul}
\usepackage{float}

\def\eq#1{equation~\ref{eq:#1}}
\def\fig#1{figure~\ref{fig:#1}}

\begin{document}
%

\title{Convolutional Polar Codes on Channels with Memory using Tensor Networks}

\author{Benjamin Bourassa${}^{1*}$,
        Maxime Tremblay${}^1$
        and David Poulin${}^{1,2}$
        \\
       ${}^1$Département de physique \& Institut quantique, Université de Sherbrooke,
     Sherbrooke, Québec, Canada J1K 2R1\\
     ${}^2$Canadian Institute for Advanced Research, Toronto, Ontario, Canada M5G 1Z8 \\
${}^*$benjamin.bourassa@usherbrooke.ca}


\maketitle

\begin{abstract}
Arikan's recursive code construction is designed to polarize a collection of memoryless channels into a set of good and a set of bad channels, and it can be efficiently decoded using successive cancellation \cite{arikan_channel_2009}. It was recently shown that the same construction also polarizes channels with memory \cite{sasoglu_polar_2016}, and a generalization of successive cancellation decoder was proposed with a complexity that scales like the third power of the channel's memory size \cite{wang_construction_2015}. In another line of work, the polar code construction was extended by replacing the block polarization kernel by a convoluted kernel \cite{ferris_convolutional_2017}. Here, we present an efficient decoding algorithm for finite-state memory channels that can be applied to polar codes and convolutional polar codes. This generalization is most effectively described using the tensor network formalism, and the manuscript presents a self-contained description of the required basic concepts. We use numerical simulations to study the performance of these algorithms for practically relevant code sizes and find that the convolutional structure outperforms the standard polar codes on a variety of channels with memory.
\end{abstract}

\IEEEpeerreviewmaketitle

\section{Introduction and Background}

\IEEEPARstart{I}{n} some communication settings, errors tend to occur in burst, which motivates the design of good error correcting protocols tailored to correlated noise models. Polar codes \cite{arikan_channel_2009} are well-known to achieve the capacity on symmetric memoryless channels and it was shown in \cite{sasoglu_polar_2016} that they also achieve polarization on channels with memory. The successive cancellation (SC) decoder can be adapted to a noise channel with a $d$-state memory at the expense of a $d^3$ increase in complexity \cite{wang_construction_2015}.

In \cite{ferris_convolutional_2017}, a generalization of polar codes was proposed which replaces the block-structured polarization kernel by a convolutional structure. The main motivation of this generalization is that it preserves the efficiency of SC decoding while significantly extending the code family. The design and analysis of convolutional polar codes are most easily formulated in terms of tensor networks. In this language, the efficiency of SC decoding of convolutional polar codes originates from a well-known ansatz in quantum many-body physics \cite{evenbly_class_2014}. In \cite{ferris_convolutional_2017} and \cite{Tremblay2018}, analytical and empirical evidences showed that convolutional polar codes outperform regular polar codes on memoryless channels both asymptotically and for finite code size, with only a small constant factor in the decoding and encoding overhead. Moreover, in \cite{morozov_efcient_nodate} an explicit description of the SC decoder for convolutional polar codes is provided and extended to the case of list decoding, showing better performance than list decoding polar codes. 

Here, we use tensor networks to describe how the convolutional polar code's SC decoder can be adapted to channels with a $d$-state memory resulting in a $d^3$ complexity increase.  For regular polar codes, this decoding algorithm coincides with the one of \cite{wang_construction_2015}. In all cases, the extension to channels with memory is particularly simple when expressed in terms of tensor networks, which illustrates the importance of this tool in coding theory. For this reason, we provide a pedagogical introduction to tensor networks in coding theory which is formulated to extend beyond polar codes. Finally, we use numerical simulations to investigate the performance of polar codes and convolutional polar codes on Gilbert-Elliott type channels, and find that in all cases, convolutional codes outperform the original construction. 

\subsection{Finite-state channels and successive cancellation decoder}

Finite-state channels consist of a class of channels where the output depends on the channel input and an internal state $s \in \mathcal{S}$ with a memory size of $|\mathcal{S}| = d$. This channel is represented by $W(Y_i, S_i|X_i, S_{i-1})$, with $X_i$ and $Y_i$ the input and output of the channel and $S_{i-1}, S_i$ the memory  states before and after the use of the channel. For a string of length $N$, the resulting channel is described by
\begin{equation}\label{fsc}
W_N(y_1^N|x_1^N,s_0) = \sum_{s_1^N} \prod_{i = 1}^NW(y_i, s_i|x_i, s_{i-1})
\end{equation}
with $s_0$ as the initial memory state. When the initial memory state is distributed with probability $P(S_0)$, we get $W_N(y_1^N|x_1^N) = \sum_{s_0} P(s_0) W_N(y_1^N|x_1^N,s_0)$.

In \cite{wang_construction_2015}, a SC trellis decoder is proposed as a generalization of the standard SC decoder, enabling an efficient decoder for polar codes on channels with memory. For a code of size $N$ with polarization transformation $G_N$,  a maximum likelihood estimator for the symbol $u_i$ conditioned on previously decoded symbols $u_1^{i-1}$ is obtained from the quantity
\begin{align}\label{synthchan}
    W^{(i)}_{N}(y_1^N| u_1^i)  &= \sum_{u_{i+1}^N}P(y_1^N | u_1^N) \\
	&= \sum_{u_{i+1}^N}W_{N}(y_1^N | u_1^NG_N).
\end{align}
Using the product structure of $W_N$ given at Eq.~\ref{fsc} and the simple form of $G_N = G_2^{\otimes n}$ with $N=2^n$,  the quantity $W^{(i)}$ can be computed in a recursive manner for polar codes \cite{wang_construction_2015}, extending the procedure known for memoryless channels \cite{arikan_channel_2009}. Above, we have expressed the channel $W^{(i)}$ as a marginal of the  probability $P(\,\cdot\,)$ of the output $y_1^N$ conditioned on the input $u_1^{i-1}$, where marginalization is taken by summing overall unknown symbols $u_{i+1}^N$. This probability is naturally expressed in terms of the channel $W_N$, which by definition is the probability of an output string given an input string. In the next section, we will explain how to express conditional and marginal probabilities  in terms of tensor networks. Once this notation is established, the decoding algorithm will be applied straightfowardly to convolutional polar codes, which illustrates the power of tensor networks.

\subsection{Tensor Networks}

Tensor networks are a computational and conceptual tool developed in the context of quantum many-body physics, see  \cite{orus_practical_2014,bridgeman_hand-waving_2017} for introductions. Similar to graphical models used in coding theory (e.g. factor graph), they are compact representations of correlated distributions involving a large number of random variables. Recently, computational problems in coding theory have been recast in the tensor network formalism \cite{ferris_tensor_2014}, a connection which has found several applications \cite{ferris_convolutional_2017,Tremblay2018,darmawan_2018,darmawan_2017,Pastawski2015,bravyi}. In the context of this paper, tensor networks can be thought of as an efficient way to represent various marginal and conditional probabilities over input bits $u_i$.

\begin{figure}[t]
\centering
\includegraphics[scale=0.5]{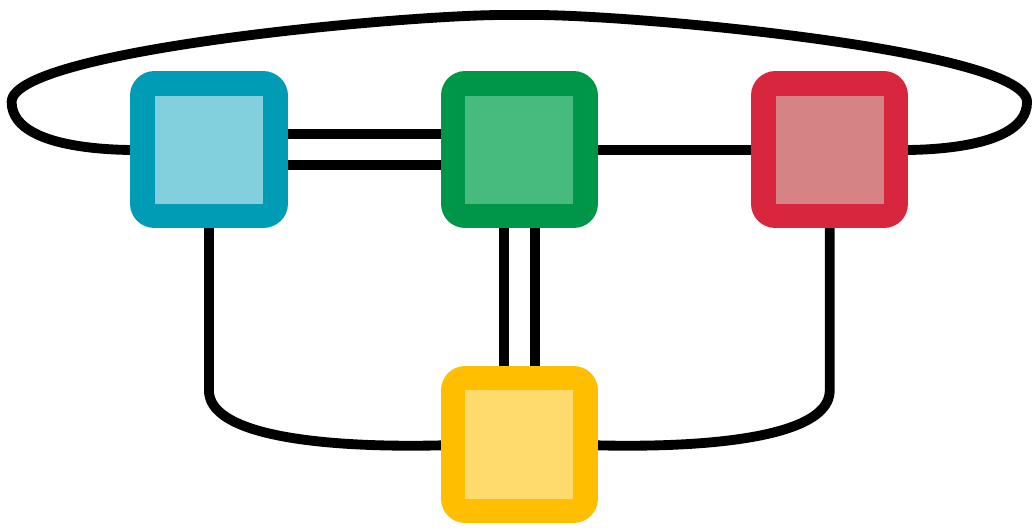}
\caption[A tensor and a tensor network]{A tensor network consisting of 4 tensors. Each vertex represents a different tensor and the rank of those tensors is given by the degree of their associated vertex. In this example all edges are contracted, so the tensor network represents a scalar.}
\label{fig:tenseur}
\end{figure}

For the present purpose, a rank-$r$ tensor is a real-valued array $A_{i_{1} i_{2} ... i_{r}}$ with $r$ indices. Each index $i_k$ has a finite range $|i_k| = \chi_k$ named bond dimension. Graphically, we represent this object by a vertex of degree $r$ where the edge $k$ is associated to the index $i_k$. The ordering of the edges should be clear from the context or specified otherwise. 

An important operation in linear algebra is  matrix multiplication. We generalize this operation by the so-called tensor contraction which consists of summing over repeated indices of same bond dimension. Given two tensors $A_{i_{1} i_{2} ... i_{r}}$ and $B_{j_{1} j_{2} ... j_{q}}$ (not necessarily of equal rank), the contraction over two indices, say $i_3$ and $j_4$, is given by
\begin{equation}\label{eq:contract}
\sum_{s = 1}^{\chi} A_{i_{1} i_{2} s ... i_{r}} B_{j_{1} j_{2} j_{3} s ... j_{q}} = C_{k_{1} ... k_{r+q-2}}, 
\end{equation}
which clearly coincides with matrix multiplication when both tensors have rank two. Graphically, this tensor contraction is represented by joining the edges from tensor $A$ and $B$ that are being contracted. For instance, the contraction of equation \ref{eq:contract} with $A$ of rank $r = 6$ and $B$ of rank $q = 4$ is represented by 
\begin{equation}\label{eq:contract2}
 \includegraphics[scale=0.5]{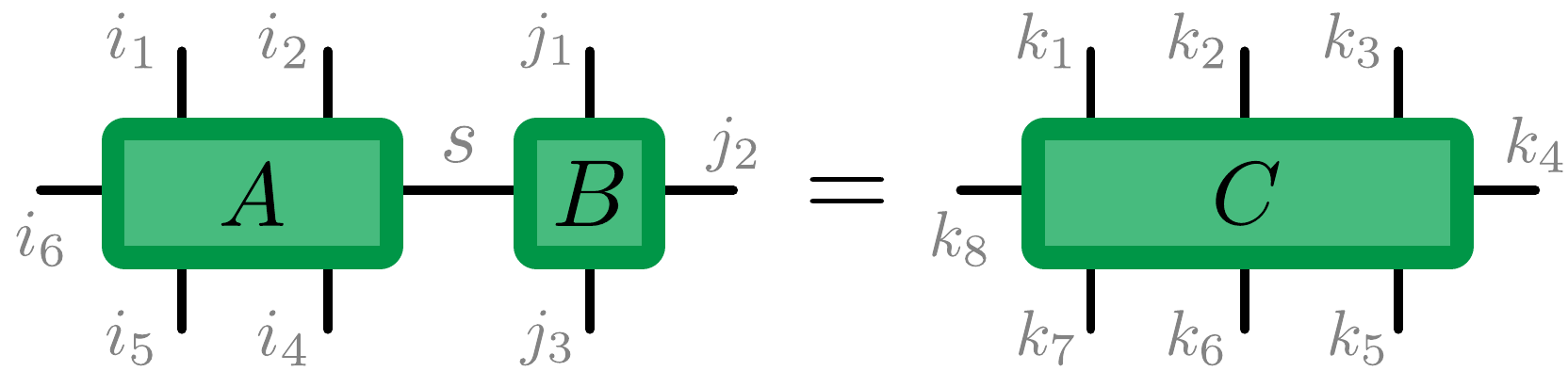}.
\end{equation}
A tensor network is a graph where each vertex represents a tensor and each edge represents a contracted index as shown in figure \ref{fig:tenseur}. The graph can also contain open edges, i.e., edges that are attached to a single vertex of tensors as in equation \ref{eq:contract2} .

Storing the entries of a tensor network requires an array of size exponential in the number of open edges. For instance, if all edges of tensors $A$ and $B$ above have bond dimension $\chi$, then $A\in \mathbb R^{\chi^r}$, $B\in \mathbb R^{\chi^q}$, and $C\in \mathbb R^{\chi^{r+q-2}}$. Thus, only tensor networks with a few open edges are of computational interest. Even for a tensor network with no open edges, the evaluation of the tensor network obtained by contracting each edge one at a time may result in tensor networks at intermediate steps with a large number of open edges, thus requiring an exponential amount of memory. Given  a tensor network, the computational task of evaluating the corresponding array -- i.e., computing all the contraction -- is generally hard (\#P-hard \cite{arad_quantum_2008}).

Despite this general hardness, some tensor networks can be efficiently evaluated. It is the case of networks with the geometry of chains or more generally trees \cite{shi_classical_2006}, which can be evaluated using standard dynamic programming methods. This is analogous to a well-known fact in coding theory, that belief propagation is exact and efficient on loop-free graphs (trees). More generally, the complexity of evaluating a tensor network scales exponentially with the network's treewidth \cite{arad_quantum_2008,markov_simulating_2008}. 

As we will explain in the next sections, some coding problems are naturally formulated in terms of tensor networks \cite{ferris_tensor_2014}. This connection facilitates the conception of new coding schemes as well as the implementation of certain decoding algorithms \cite{Tremblay2018,darmawan_2018,darmawan_2017,Pastawski2015,bravyi}. In particular, previous work connecting tensor networks to polar codes with i.i.d. channels \cite{ferris_convolutional_2017} naturally extends to channels with memory.

\section{Decoding algorithm}

\subsection{Probability density as a tensor network}

Before describing the details of the decoding process, we explain how tensor networks can be used to represent conditional and marginal probabilities involving many variables. We specialize to binary random variables for sake of clarity, but the techniques naturally extend to larger alphabets. Thus, consider a probability distribution of the $N$ input bits $P(U_1^N)$. This object is a real and positive array of dimension $2^N$, but it can equivalently be seen as the entries of a rank-$N$ tensor with each index having bond dimension 2,
\begin{equation}
  \includegraphics[scale=0.7]{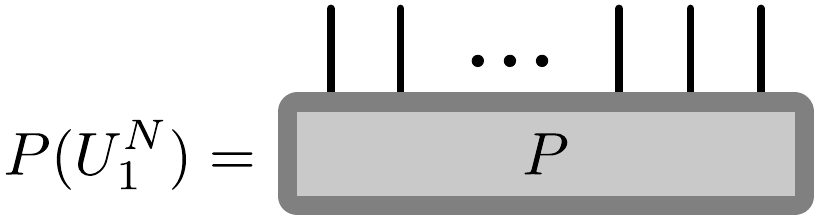}.
\end{equation}

In the context of the decoding process, the object of interest is the probability of the input $U_1^N$ conditioned on a fixed output $y_1^N$, and Bayes's rule gives 
\begin{align}
P(U_1^N|y_1^N) &= P(y_1^N|U_1^N) \frac{P(U_1^N)}{P(y_1^N)} \\
& = W_N(y_1^N|U_1^NG_N)  \frac{P(U_1^N)}{P(y_1^N)} \\
&\propto W_N(y_1^N|U_1^NG_N)
\end{align}
where in the last line we have used the fact that $y_1^N$ is fixed, so $P(y_1^N)$ is a constant, and we have assumed a uniform prior over inputs $P(U_1^N)$.
Maximum likelihood decoding consists in maximizing  the probability $P(U_1^N|y_1^N)$ overall inputs $U_1^N$, an exponentially large set. Because of this exponential growth, maximum likelihood decoding is generally intractable. 

In the context of polar codes, the use of a SC decoder replaces this global optimization procedure by sequence of bitwise optimizations. At the $i$-th step of decoding, the input bits $u_1^{i-1}$ are known and the input bits $u_{i+1}^N$ are ignored, so summed over, resulting in the quantity $P(u_i|y_1^N,u_1^{i-1}) = \sum_{u_{i+1}^N}P(u_1^N|y_1^N)$.  The $i$-th bit is decoded to the value $\hat u_i$ which optimizes this conditional probability, i.e. $\hat u_i = \arg \max_{u_i} P(u_i|y_1^N,u_1^{i-1})$, which is equivalent to optimizing equation \ref{synthchan}.

To express this procedure as a tensor network, we define three rank-one tensors
\begin{equation}
\includegraphics[scale=0.7]{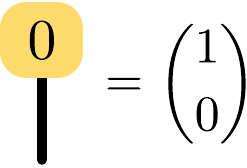}, \qquad  \includegraphics[scale=0.7]{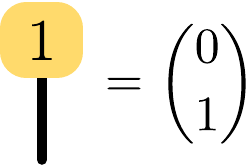} \quad{\rm and} \quad  \includegraphics[scale=0.7]{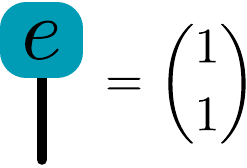}.
\label{eq:dist}
\end{equation}
The first two tensors represent a variable with a fixed value 0 or 1 respectively, while the third tensor, when contracted to another edge, has the effect of summing over the values of the associated variable. Then, the $i$-th step of SC maximum likelihood decoding is expressed graphically as
\begin{equation}\label{scd}
\includegraphics[scale=0.7]{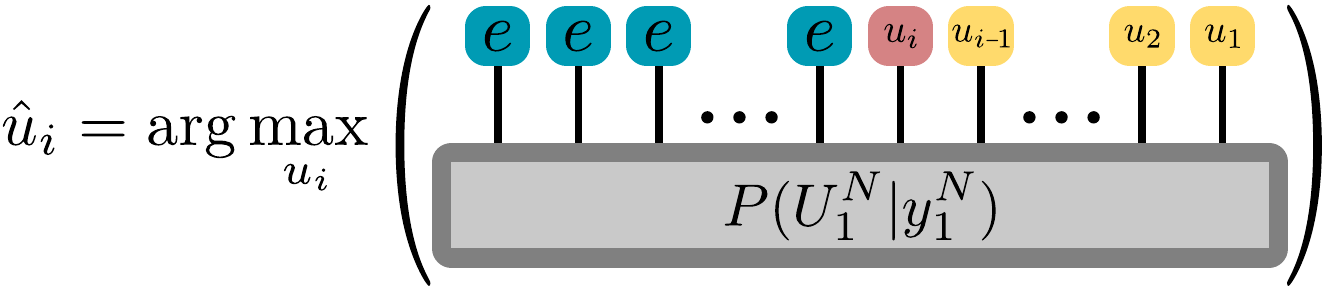}.
\end{equation}
Indeed, the yellow tensors to the right fix the value of previously decoded bits $u_1^{i-1}$ while the left tensors to the left sum over the latter bits $u_{i+1}^N$, reproducing equation \ref{synthchan}.

Since $P(U_1^N|y_1^N)$ is proportional to the likelihood $W_N(Y_1^N|U_1^NG_N)$, we can decompose $P$ in the following way,
\begin{equation}\label{tn}
\includegraphics[scale=0.7]{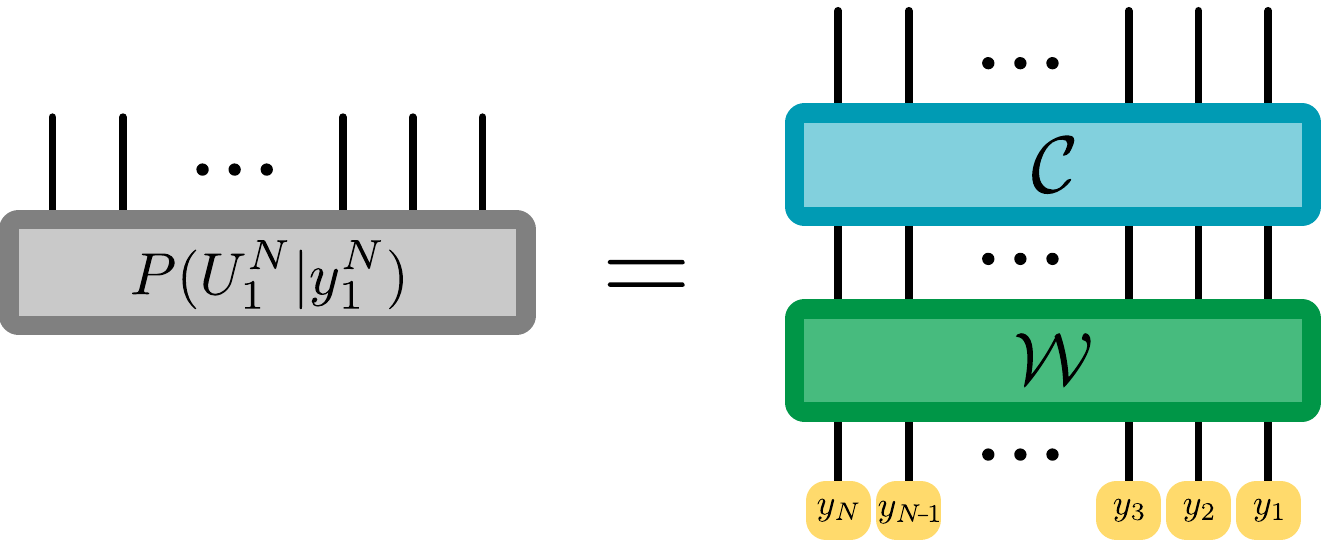}
\end{equation}  
where $\mathcal{C}$ is a tensor representation of the encoding circuit which realizes the transformation $G_N$ (see section \ref{ssec:dec_TN}), and $\mathcal{W}$ is a description of the noise model in terms of a tensor network (see section \ref{ssec:channel_TN}). Equation \ref{tn} is very general and could be conceptually applied to any linear  or non-linear code. Contracting this tensor is in general inefficient because it does not have a bounded treewidth. We will see in the next section that SC decoding as defined in equation \ref{scd} results in a tensor network of constant treewidth when using polar codes and convolution polar codes under finite-state channels, yielding an efficient decoder that generalizes the decoder used in \cite{ferris_convolutional_2017} and \cite{wang_construction_2015}.
 
\subsection{Channel model as a tensor network}
\label{ssec:channel_TN}

A binary discrete memoryless channel $W$ is completely described by a stochastic matrix $W(Y|X)$ where $X$ denotes the input and $Y$ the output. This is a rank-2 tensor with one edge representing the input $X$ and the other representing the output $Y$. The i.i.d. channel acting on an $N$-bit sequence $X_1^N$ can be represented as a product of $N$ independent tensors as follows
\begin{equation}
  \includegraphics[scale=0.5]{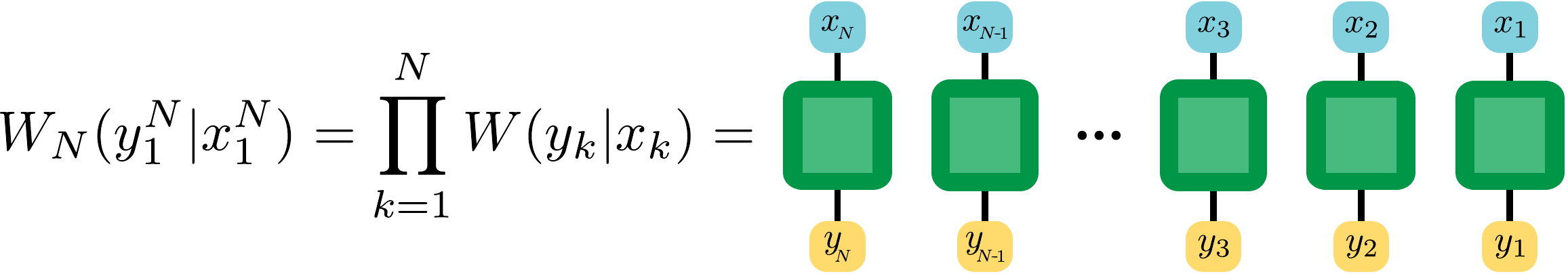}.
  \label{eq:iid}
\end{equation}

For finite-state channels, we define a rank-four tensor
\begin{equation}
\includegraphics[scale=0.5]{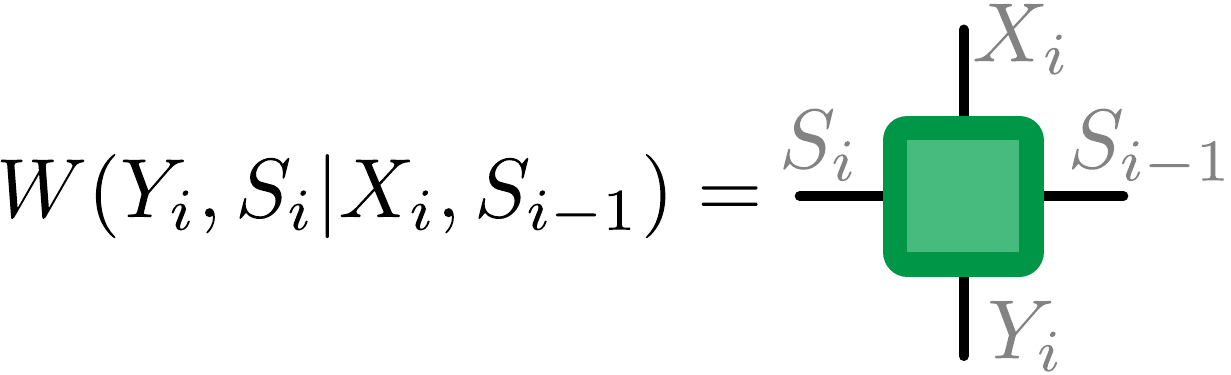},
\end{equation}
with horizontal edges of bond dimension $d$, the size of the channel's memory, while the top vertical edges have the dimension of the input alphabet $X$ and lower vertical edges have the dimension of the output alphabet $Y$.
The channel with memory of equation \ref{fsc} is obtained by summing over the memory states, so it is represented by the tensor network 
\begin{equation}
\includegraphics[scale=0.5]{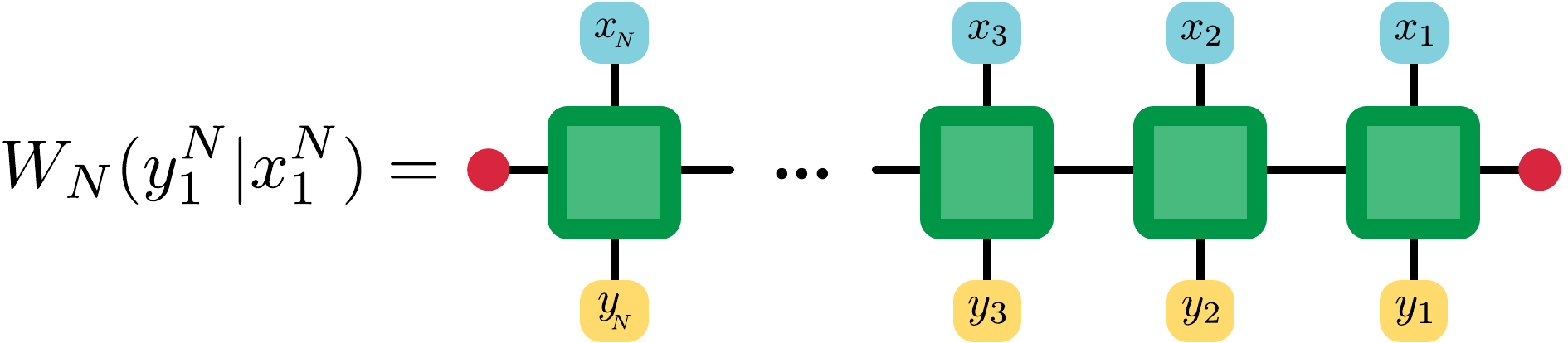}.
\label{eq:fsc_tn}
\end{equation}
The rightmost rank-1 tensor corresponds to the distribution of the initial state $P(S_0)$, whereas the leftmost  tensor is an all-1 vector which has the effect of summing over the final memory state $S_N$. This chain structure is quite familiar in condensed matter physics where it goes under the name {\em matrix product operator} (MPO) \cite{mpo}.  

\subsection{Decoding as tensor network contraction} \label{ssec:dec_TN}

        \begin{figure}
            \centering
    		\subfloat[]{
        		\includegraphics[width=3in]{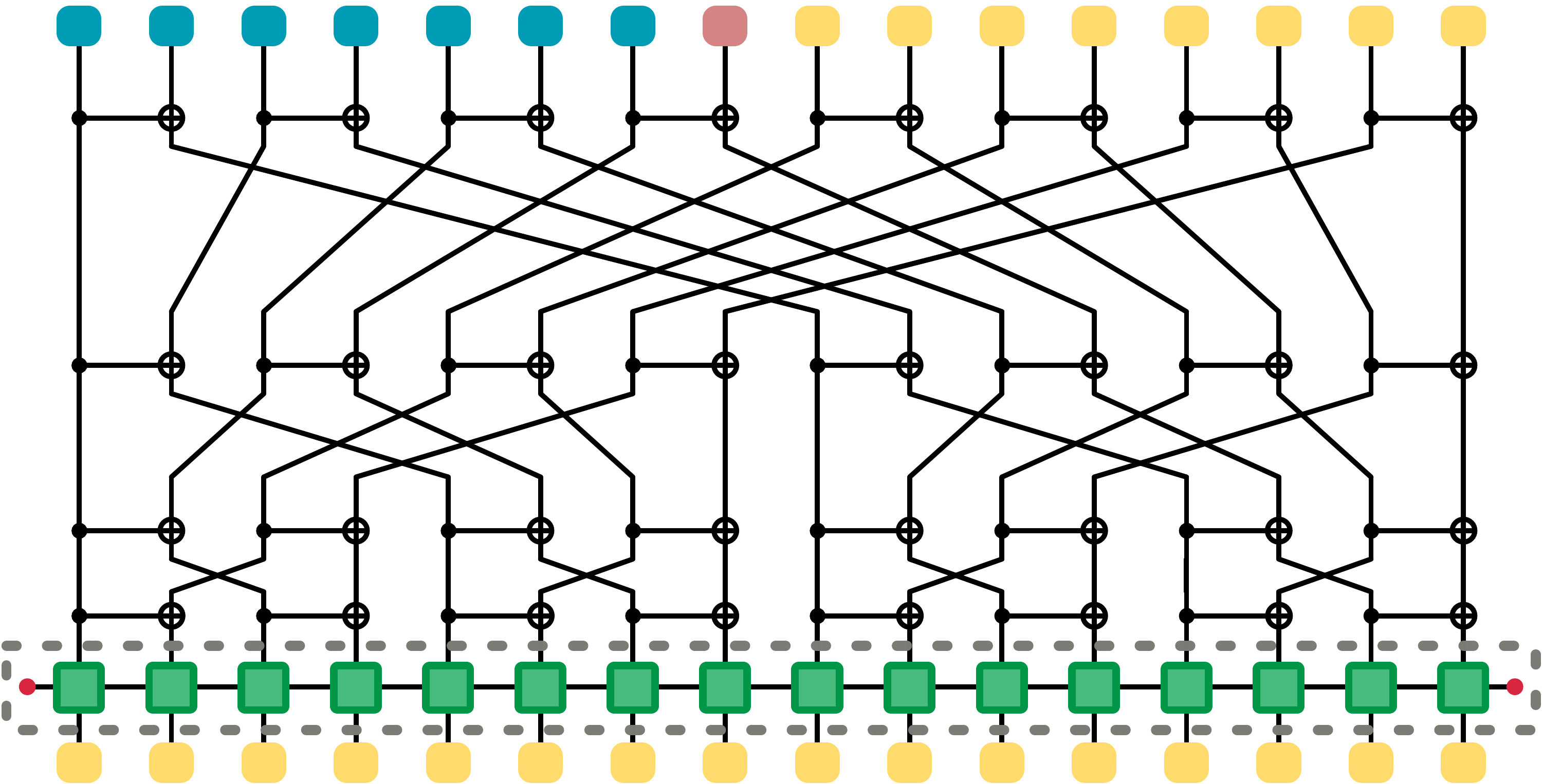}
        		\label{fig:tn_pc}}
    		\quad
    		\subfloat[]{
        		\includegraphics[width=3in]{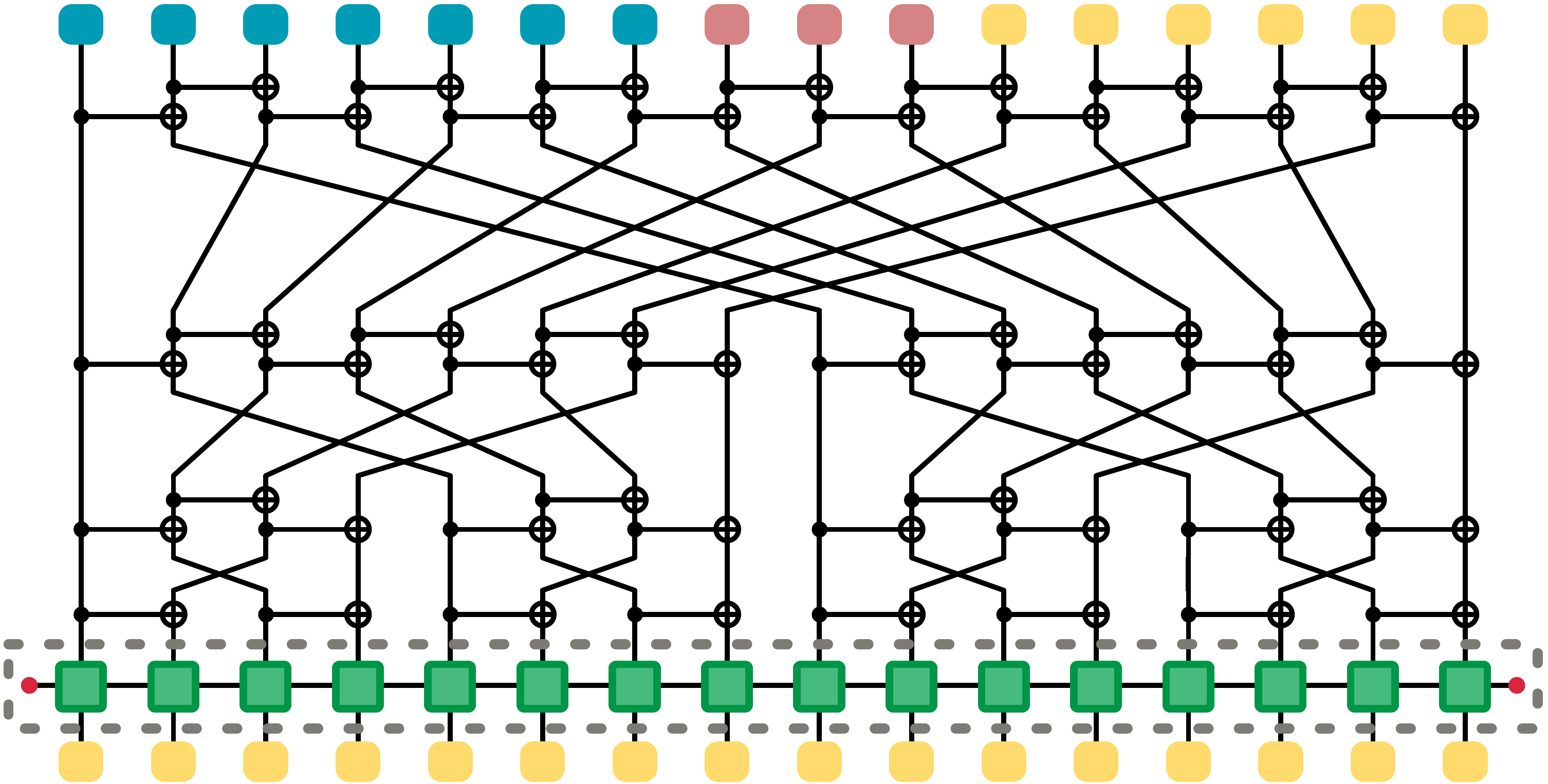}
        		\label{fig:tn_cpc}}
        	\quad
            \subfloat[]{
                \includegraphics[width=3in]{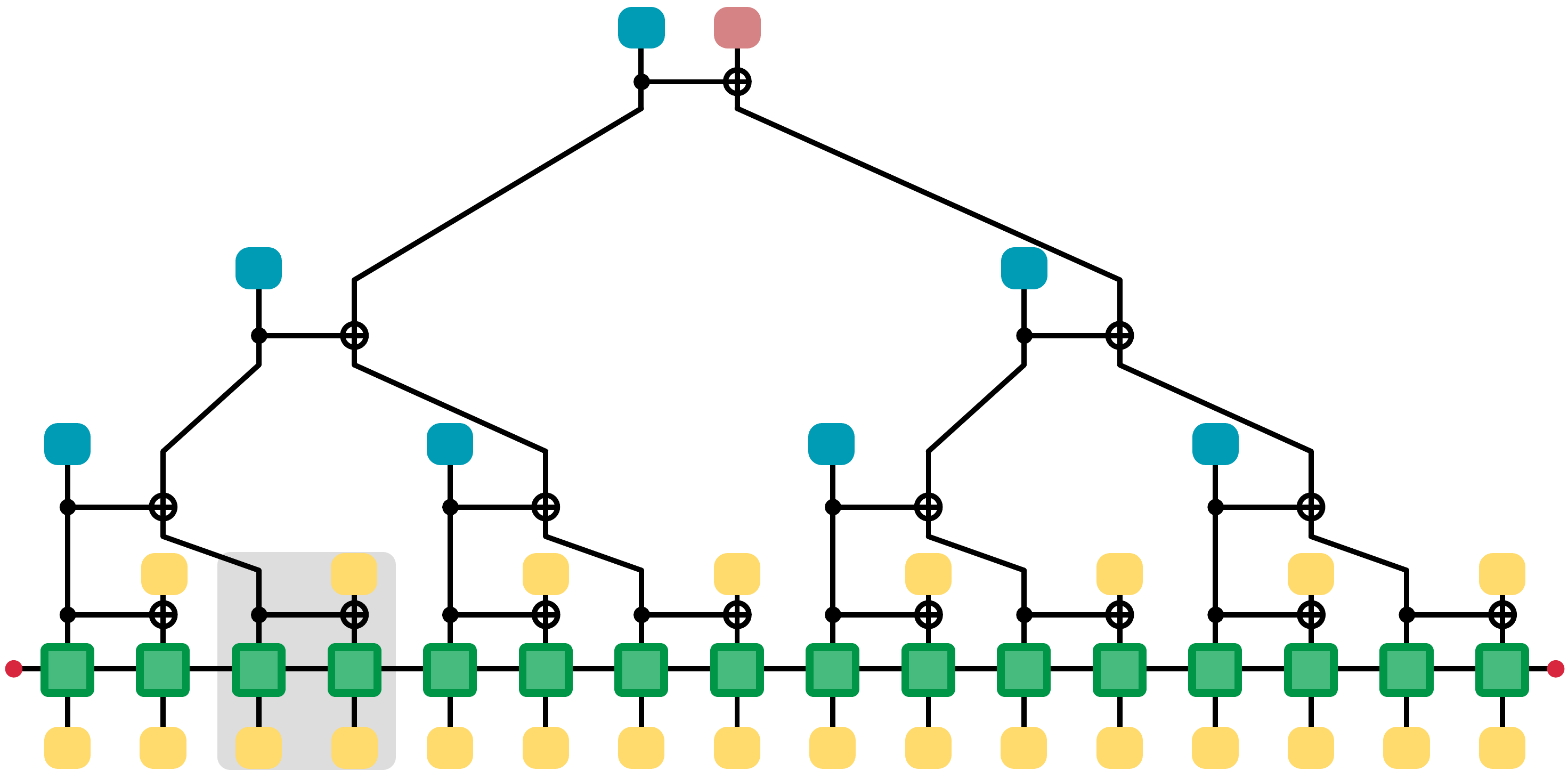}
                \label{fig:simp_pc}}
            \quad
            \subfloat[]{
                \includegraphics[width=3in]{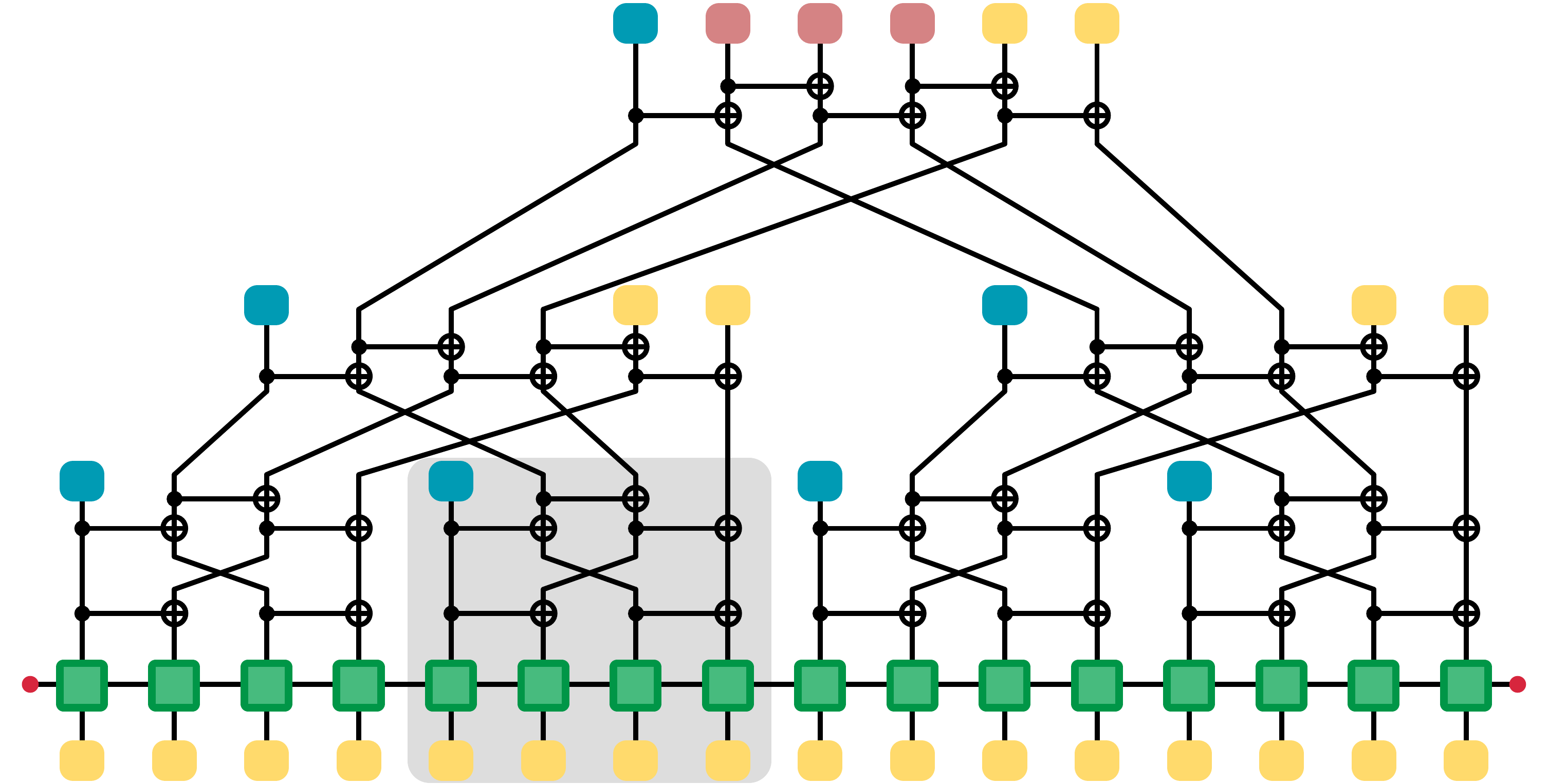}
                \label{fig:sim_cpc}}
            \quad
            \subfloat[]{
                \includegraphics[width=3in]{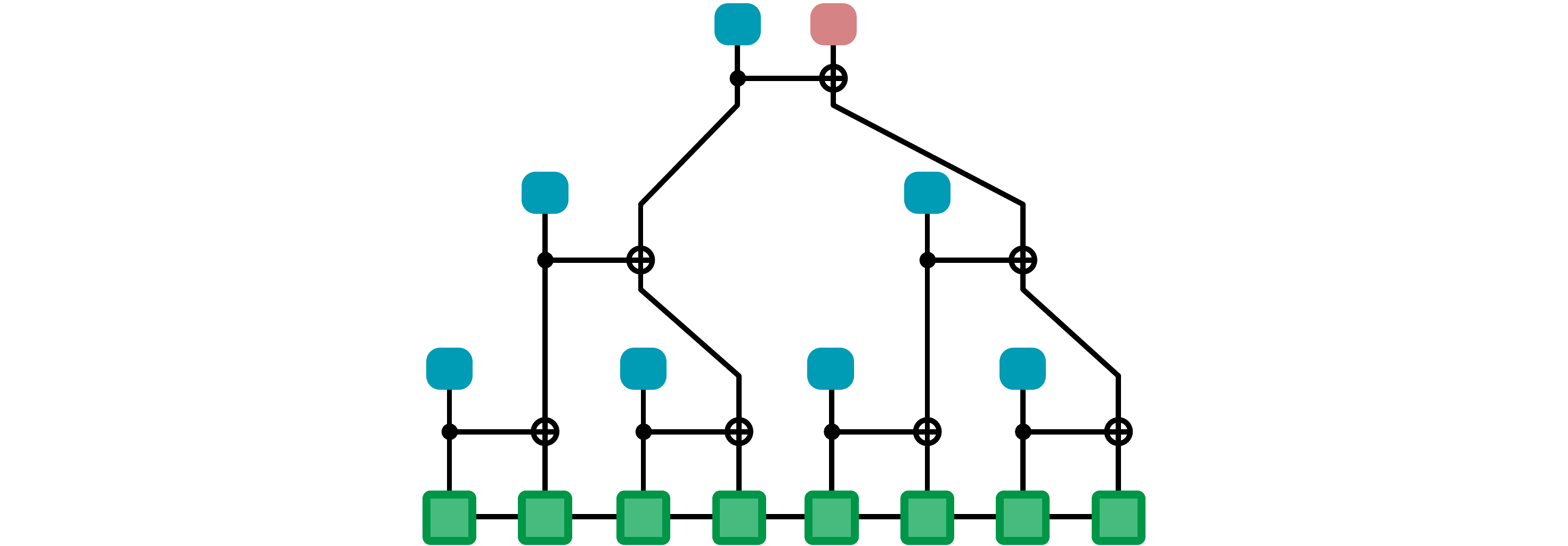}
                \label{fig:red_pc}}
            \quad
            \subfloat[]{
                \includegraphics[width=3in]{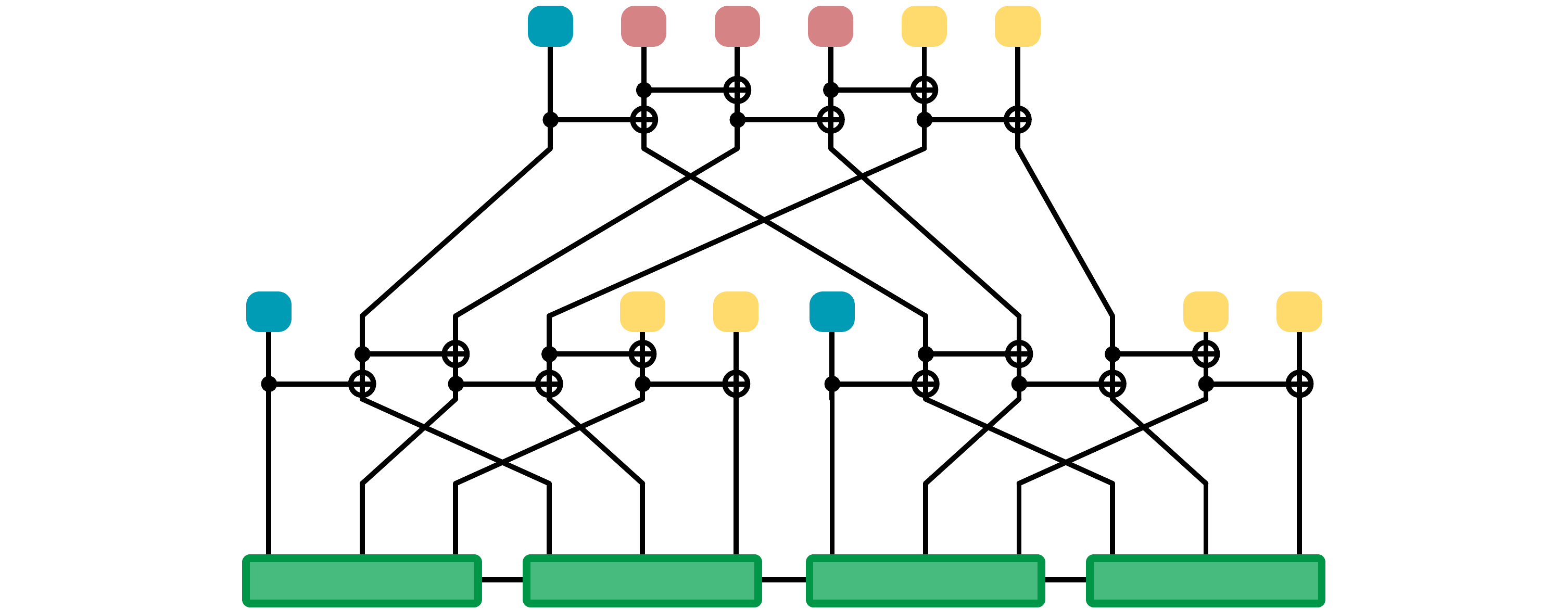}
                \label{fig:red_cpc}}                
            \caption{Subfigure \textbf{(a)} and \textbf{(b)} represent the tensor networks corresponding to one step of the SC decoder for a polar code and a convolutional polar code respectively. The channel is one with memory described at equation \ref{eq:fsc_tn}, but could be replaced by an i.i.d. channel by replacing the tensor network in the dotted region by equation \ref{eq:iid}. Subfigure \textbf{(c)} and \textbf{(d)} show the resulting tensor networks after the simplification procedure given by equation \ref{simplify} for polar codes and convolutional polar codes respectively. The gray region represents the initial contraction of the decoding algorithm. The blue rank-one tensors represent the $e$ tensor of equation \ref{eq:dist}, whereas the yellow one represent known bits values and the red tensors on the top of the circuit represent the bits being decoded. Subfigure \textbf{(e)} and \textbf{(f)} represent the reduce tensor networks resulting of the initial contraction given by equation \ref{eq:init_contraction_pc} for polar codes and \ref{eq:init_contraction_cpc} for convolutional polar codes.}
            \label{fig:simplification}
        \end{figure}

As mentioned above, the encoding transformation $\mathcal{C}$ is a tensor of rank $2N$. We can think of $N$ of the edges as input bits (including frozen bits) and the other $N$ edges as output bits. Any such encoding transformation can be decomposed in terms of elementary transformations by an encoding circuit, which corresponds to factoring a high-rank tensor network into a tensor network composed of constant rank tensors. For polar codes and convolutional polar codes, the encoding tensor network is factorised into a network of rank-4 tensors. The primitive of this factorization is the standard polarization kernel, the controlled-not. The corresponding rank-4 tensor takes value 1 if $c =a$ and $d = a \oplus b$ and takes value 0 otherwise. In other words, we can think of this tensor as having two input indices ($a$ and $b$) and two output indices ($c$ and $d$), and its value is the indicator function of the controlled-not transformation. Graphically, we represent this tensor as
\begin{equation}
\includegraphics[scale=0.8]{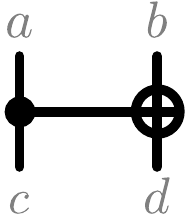}.
\label{eq:CNOT}
\end{equation}
Alternatively, the action of this tensor on a probability distribution $P(a,b)\in \mathbb R^4$ is $ P(a,b) \rightarrow Q(c,d) = P(a,a\oplus b)$. Graphically, it is represented by 
\begin{equation}
\includegraphics[scale=0.8]{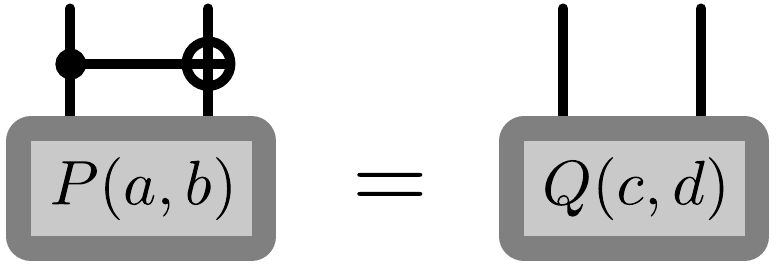}.
\label{eq:cnot_contract}
\end{equation}

At step $i$, SC decoding consists in evaluating the tensor network of figure \ref{fig:tn_pc} for polar codes and figure \ref{fig:tn_cpc} for convolutional polar codes. As in equation \ref{scd}, the yellow tensors to the right fix the value of previously decoded bits while the blue tensors to the left have the effect of summing over the value of the later bits. As can be seen in figure \ref{fig:tn_cpc}, sequences of 3 bits are decoded simultaneously on convolutional polar codes (a consequence of the graph treewidth which is larger than standard polar codes). A priori, both of these tensor networks have a treewidth that scales with $N$, so they cannot be contracted efficiently. However, there exist algebraic relations between the elementary tensors which lead to important simplifications. These identities are 
\begin{equation}\label{simplify}
\includegraphics[scale=0.5]{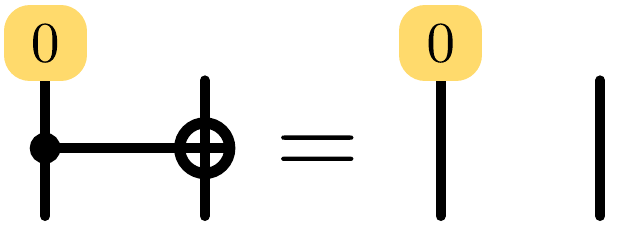} ,\qquad \includegraphics[scale=0.5]{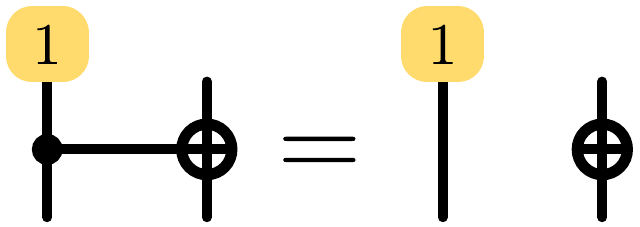}, \qquad \includegraphics[scale=0.5]{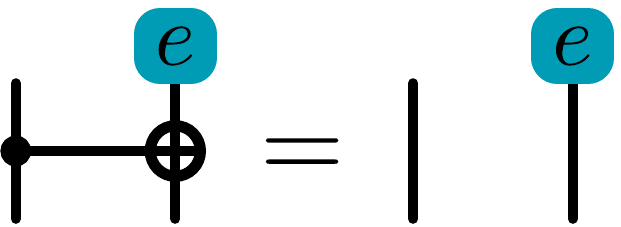}
\end{equation}
and should have a clear intuitive meaning. Applying these algebraic identities to the tensor network of figure \ref{fig:tn_pc} and  \ref{fig:tn_cpc} result in tensor networks that have the topology of trees, where the bottom leaves are connected to a chain, i.e. figures \ref{fig:simp_pc} and \ref{fig:sim_cpc} respectively.

The decoding task then consists of contracting these tensor networks. While it is fairly intuitive that a tree tensor network is contractible with complexity $O(N)$, when contracting from bottom to top, the presence of a chain at the bottom of the network makes matters slightly more complex. The contraction can nonetheless be performed efficiently. One way to see this is to note that the resulting graphs have constant treewidth. A more informative explanation uses graphical identities that we now explain. 

Begin with the standard polar code in figure \ref{fig:simp_pc}. The contraction algorithm starts from the bottom leaves where it performs the following {\em initial contraction} 
\begin{equation}
\includegraphics[scale=0.5]{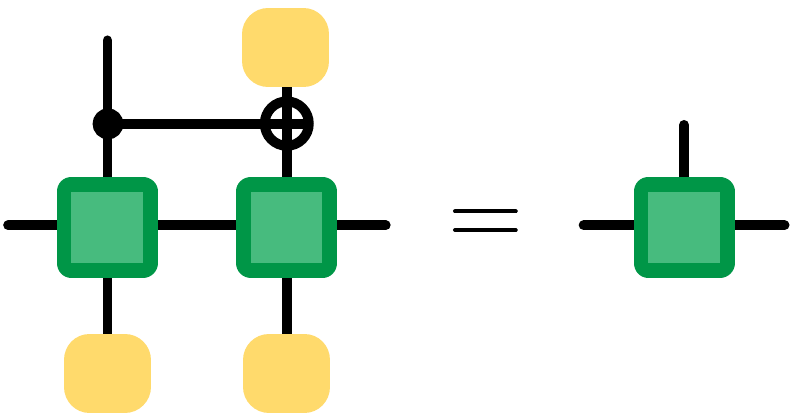}.
\label{eq:init_contraction_pc}
\end{equation}
An example is illustrated by a shaded grey region in figure \ref{fig:simp_pc}. This equation only conveys the graphical structure of the tensor network, but this is sufficient to demonstrate that each step is efficient. The important information to take away from this equation is that the tensor network to the left -- which combines 3 rank-one tensors and 3 rank-four tensors -- produces a rank-tree tensor shown on the right. The value of this tensor is obtained by summing over the contracted indices and clearly this operation requires a number of operations which is {\em independent of $N$}. It can easily be seen that the complexity of this operation scales like $d^3$ because it involves 3 bond-dimension $d$ edges (as well as some bond-dimension 2 edges). Once all the initial contractions have been performed on the leaves of the tree, we obtained a reduced tensor network as shown in figures \ref{fig:red_pc}. The following {\em base contraction} 
\begin{equation}
\includegraphics[scale=0.5]{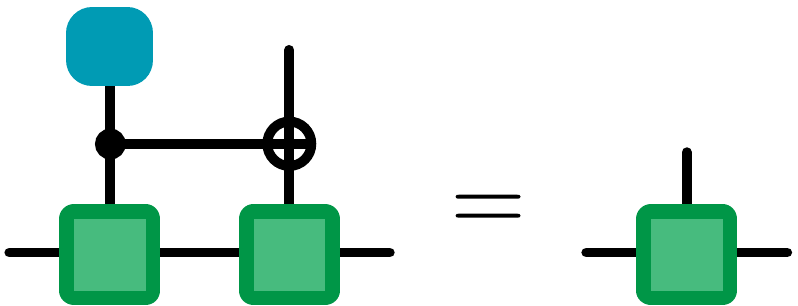}
\label{eq:contraction_pc}
\end{equation}
is then used at the leaves of the resulting tensor network, removing another layer of leaves. The base contraction is used on each layer until the root of the tree is reached.

For convolutional polar codes, the global procedure is unchanged, but instead of decoding a single bit at a time, we must decode sequences of 3 bits at a time. Hence, the initial contraction is
\begin{equation}
\includegraphics[scale=0.5]{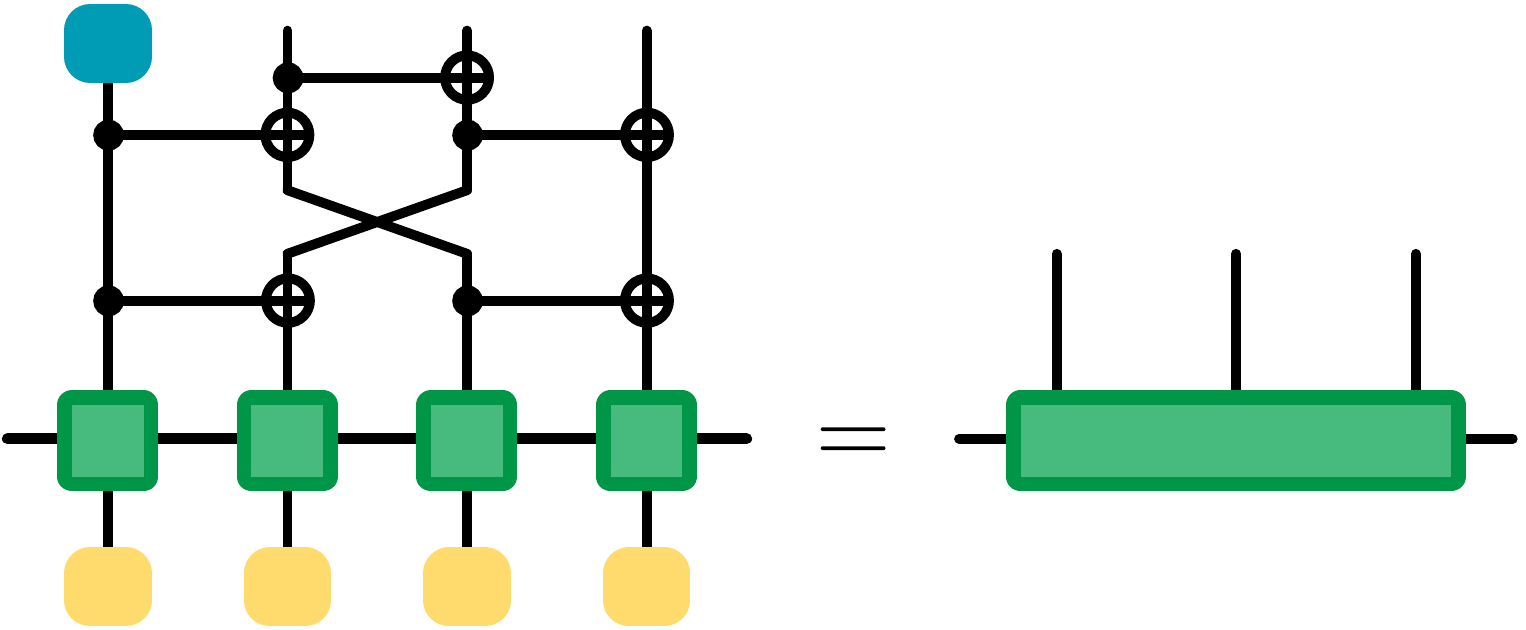},
\label{eq:init_contraction_cpc}
\end{equation}
and illustrated in the shaded grey region of figure \ref{fig:sim_cpc}. Once initial contractions are applied to all leaves of the tensor network, we obtain the tensor network illustrated at figures \ref{fig:red_cpc}.  This tensor network is contracted by a sequence of base contraction 
\begin{equation}
\includegraphics[scale=0.5]{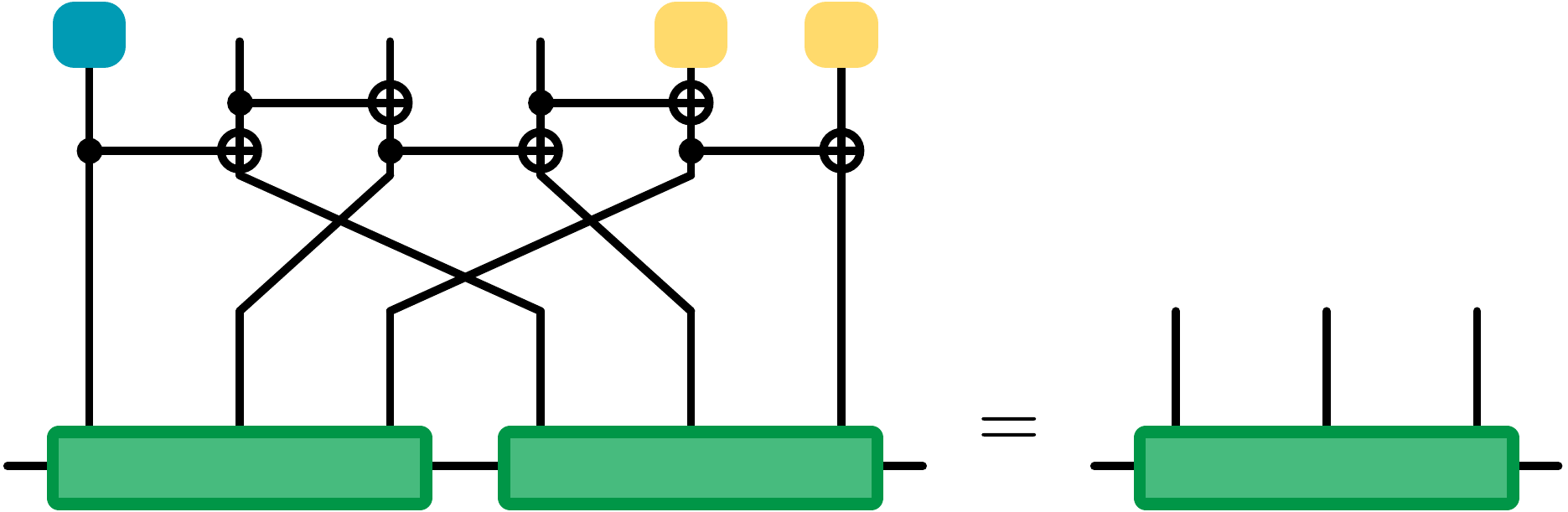}.
\label{eq:contraction_cpc}
\end{equation}
While both initial and base contractions are more complex than for standard polar codes, they again only involve a constant number of contractions, so the complexity of each operation is independent of the number of bits $N$.

In all cases, these graphic identities demonstrate that the contraction of tensor networks from figures \ref{fig:simp_pc} and \ref{fig:sim_cpc} can be broken into elementary steps which each involves constant-rank tensor networks, and hence have a complexity independent of $N$. The overall complexity for contracting the networks presented in \fig{simplification} is $O(d^3N)$. This procedure needs to be performed $N$ times to decode all $N$ bits, but it is possible to recycle partial contractions from previously decoded bits to produce a global decoding complexity of $O(d^3N\log{N})$.

\section{Numerical Results}

The encoding circuit $G_N$ does not entirely specify the code: one must in addition specify the frozen-bit locations $\mathcal F$. In our simulations, we used the following heuristic code construction. For each position $i$, we evaluate the tensor network of \fig{tn_pc} for polar codes or \fig{tn_cpc} for convolutional polar codes with all bits $U_1^{i-1}$ fixed to 0, bit $U_i$ fixed to 1, and all positions to the left of $i$ fixed to the rank-1 tensor $e$. This tensor evaluates to a value $E(u_i)$, that is, the probability of being the first undetected error, and we freeze the $n - k$ bits with the largest value of $E(u_i)$. As observed in \cite{ferris_convolutional_2017}, the quantity $E(u_i)$ serves as a good proxy for channel selection.

We simulated three types of decoding algorithms and corresponding frozen bit sets. In the first set of simulations, we use the memoryless decoding algorithm and code construction, {\em assuming a memoryless channel}. More precisely, the code construction and decoding algorithm used an i.i.d. channel with error rate equal to the average error rate of the channel with memory, but the noise in the simulation is generated from a channel with memory. This set of simulation serves as a base case to see how much is gained by taking into account known memory effects in the channel. 

The second set of simulations also assumes an i.i.d. channel with the right average error probability for code construction and decoding, but errors in the simulations are generated according to an interleaving procedure. This set of simulation is used to compare the performance of our decoder to a coding procedure where correlations are ``scrambled" using an interleaving procedure, and we will denote these simulations with the prefix {\em int}. 

The third set of simulations exploit the full correlated structure of the channel both in the code design and decoding algorithm, as described in the previous section. We use the prefix {\em corr} to denote these simulations. Finally, all three simulation sets include polar codes \emph{pc} and convolutional polar codes \emph{cpc}.

Our simulations use the Gilbert-Elliott channel\footnote{See Appendix A for details of the noise model in terms of tensor networks.} with $h_{G} = 0$ so that state $G$ is completely noiseless and $h_B = h$. This special case is usually referred to as the Gilbert channel. The error burst length $\ell_{B}$ corresponds to the length of a consecutive stay in state $B$. It has a geometric distribution with average $\langle \ell_{B} \rangle = {1}/{P_{B \rightarrow G}}$. Another channel characteristic is the \emph{good-to-bad ratio}
$$
\rho = \frac{P(G)}{P(B)} = \frac{P_{B \rightarrow G}}{P_{G \rightarrow B}}.
$$ 
The limits $\rho\rightarrow \infty $ and $\rho \rightarrow 0$ correspond respectively to a noiseless channel and ${\rm BSC}(h)$. 

\begin{figure}
\centering
\includegraphics[scale=1]{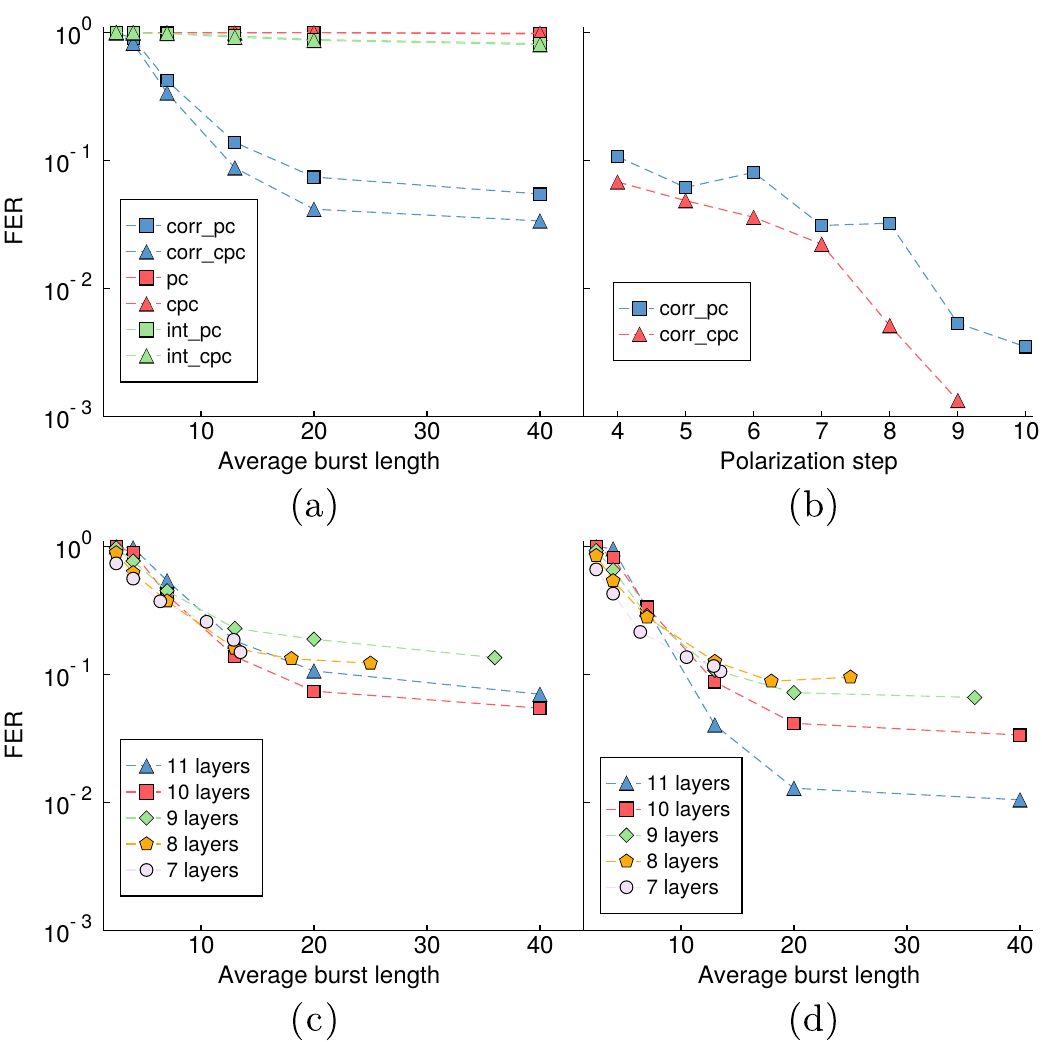}
\caption{Frame error rate (FER) for various codes, decoders and Gilbert channel parameters. (a) Varying the channel's average burst length for fixed $h=0.9$, $\rho = 5$, and average error $P(B)h = 0.15$, the exact parameters can be found in Table \ref{noise_param}. Squares show polar code \textit{pc} while triangles show convolutional polar code \textit{cpc}, both with 10 polarization steps and rate $\frac 12$. The regular decoder (red) results in high FER, using an interleaver (green) to mitigate correlations offers little improvement, while using the correlated decoder (blue) leads to a substantial reduction of the FER. b) Correlated decoder used on polar code (pc, blue squares) and convolutional polar codes (cpc, red triangles) on a channel with parameters $h = 0.9$, $P_{B\rightarrow G} = 0.05$ and $P_{G\rightarrow B} = 0.01$ with a code rate of $\frac 13$ as a function of the number of polarization steps. c) \& d) As in a) for various polarization steps and using the correlated decoder for c) polar code and d) convolutional polar code.}
\label{fig:results}
\end{figure}

Our main results are presented at \fig{results}, showing the frame error rate (FER) for various codes, decoding algorithms and channel parameters. We observe in \fig{results}a that, for both polar and convolutional polar codes, the use of an interleaver to mitigate burst errors offers only a mild improvement of performance compared to the base case where correlations are simply ignored, but that using a decoder that fully exploit the correlations offers a significant improvement. This is true for a variety of channel parameters, and as expected the improvement is reduced when the burst length becomes close to 1, in which case we recover an i.i.d. channel.

\begin{table}
\begin{center}
\begin{tabular}{|c|cccccc|}
  \hline
  $\langle \ell_{B} \rangle$ & 2.4 & 4 & 7 & 13 & 20 & 40 \\
  \hline
  $P_{B \rightarrow G}$ & 0.400 & 0.250 & 0.145 & 0.750 & 0.050 & 0.025 \\
  \hline
  $P_{G \rightarrow B}$ & 0.080 & 0.050 & 0.029 & 0.015 & 0.010 & 0.005 \\
  \hline
  $C$ & 0.507 & 0.602 & 0.696 & 0.780 & 0.817 & 0.861 \\
  \hline
\end{tabular}
\end{center}
\caption{Channel parameters for the simulations of \fig{results}\rm a, c \& d.}
\label{noise_param}
\end{table}

Figure \ref{fig:results}b presents a direct comparison of polar and convolutional polar codes under correlated decoders. In both cases, we observe a suppression of the FER as a function of the number of polarization steps, but the convolutional polar code achieves a larger suppression rate. This is consistent with the findings of \cite{ferris_convolutional_2017} in the context of i.i.d. channels and is suggestive that convolutional polar codes may achieve a larger error exponent on a wide variety of channels.

Finally, \fig{results}c and \fig{results}d present the performance of the polar codes and convolutional polar codes with varying number of polarization steps as a function of the channel's average burst length. Again, the performance of the convolutional polar codes are systematically better than those of the polar codes. In particular, for the convolutional polar codes, we observe a threshold behavior: there is a critical average burst length $\langle \ell_B\rangle^*\approx 10$ beyond which the FER decreases with the number of polarization steps. This threshold behavior is not observed for the polar codes. This observation is suggestive of two possible hypotheses: 1) In this parameter regime, the error probability will asymptotically vanish for the convolutional polar codes but not for the polar codes, i.e. polar codes cannot polarize these channels while convolutional polar codes can. 2) The finite size effects are much more prominent in the polar codes than in the convolutional polar codes. Both of these hypotheses are interesting and warrant further investigation of convolutional polar codes.

\section{Conclusion}
Tensor network is a natural language in which to formulate the decoding problem. In this language, the efficiency of SC decoding of polar codes follows immediately from the tree structure of the corresponding network. The generalization to convolutional polar codes is also efficient because the corresponding tensor network has constant treewidth. What we have shown here is that decoding a channel with memory increases the treewidth of the resulting tensor network only by a constant amount, so the decoding complexity is only affected by a constant factor. The simplicity of these results illustrates the use of tensor networks in coding theory.

Our numerical simulations confirm that a substantial improvement (up to 12dB increase noise suppression in our simulations) is obtained when taking channel correlations into account during the decoding process and code construction. Moreover, we found that convolutional polar codes outperform regular polar codes (up to 5dB increase noise suppression in our simulations) on a wide variety of channels with memory, extending the findings of \cite{ferris_convolutional_2017} beyond the i.i.d. case. In particular, our simulations suggest the existence of a range of channel parameters where polar codes fail at polarizing while convolutional polar codes succeed, an observation which warrants further investigation into convolutional polar codes.

\appendices
\section{Details of the noise model used in the simulation}

If a channel is characterized by a state process independent of the output, it can be factored as $W(y_n,s_n|x_n, s_{n-1}) = p(y_n|x_n,s_{n-1})q(s_n|x_n,s_{n-1})$ so that the state evolves stochastically through a Markov chain describing transition between the various states of the channel. We will assume that the state process is also independent of the input so that $q(s_n|x_n,s_{n-1}) = q(s_n|s_{n-1})$. This special case refers to finite-state Markov channel. The complete description of these channels is given by 
\begin{equation}\label{markov}
W_N(y_1^N|x_1^N,s_0) = \sum_{s_1^N} \prod_{n = 1}^Np(y_n|x_n,s_{n-1})q(s_n|s_{n-1}).
\end{equation}

For sake of illustration, we will consider the Gilbert-Elliott channel \cite{gilbert_capacity_1960} which has two internal states, labeled $G$ for \emph{Good} and $B$ for \emph{Bad}. When the channel is in state $G$, the transmitted bits are affected by a Binary Symmetric Channel ${\rm BSC}(h_{G})$ and when the channel is in state $B$, the transmitted bits are affected by ${\rm BSC}(h_{B})$. The good channel is less noisy than the bad channel, meaning $h_{G} < h_{B}$. This channel is known to be in the class of indecomposable finite-state Markov channel. 

In the case of finite-state Markov channel, the equation \ref{markov} can be written in the following way,
\begin{equation}
W_N(y_1^N|x_1^N,s_0) = \sum_{s_1^N}\prod_{n=1}^N \sum_{c} q(s_n|c) \delta_{c,s_{n-1}}p(y_n|x_n,c).
\end{equation}
If we do the following assignation,
\begin{equation}
\includegraphics[scale=0.5]{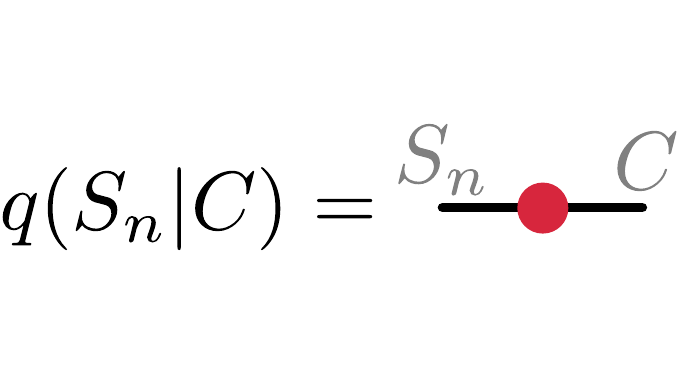} \; , \qquad \includegraphics[scale=0.5]{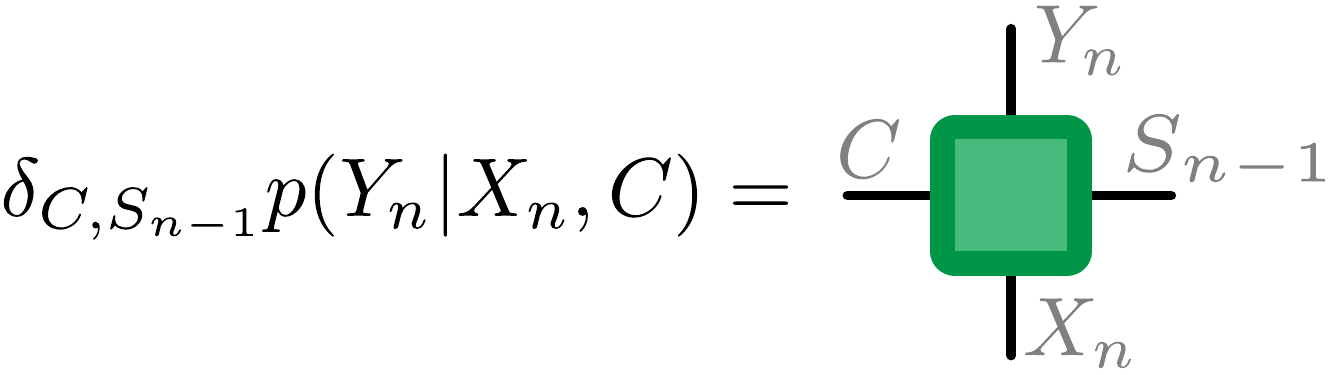}
\end{equation}
we find that equation \ref{markov} can be expressed with the following tensor network,
\begin{equation}
\includegraphics[scale=0.5]{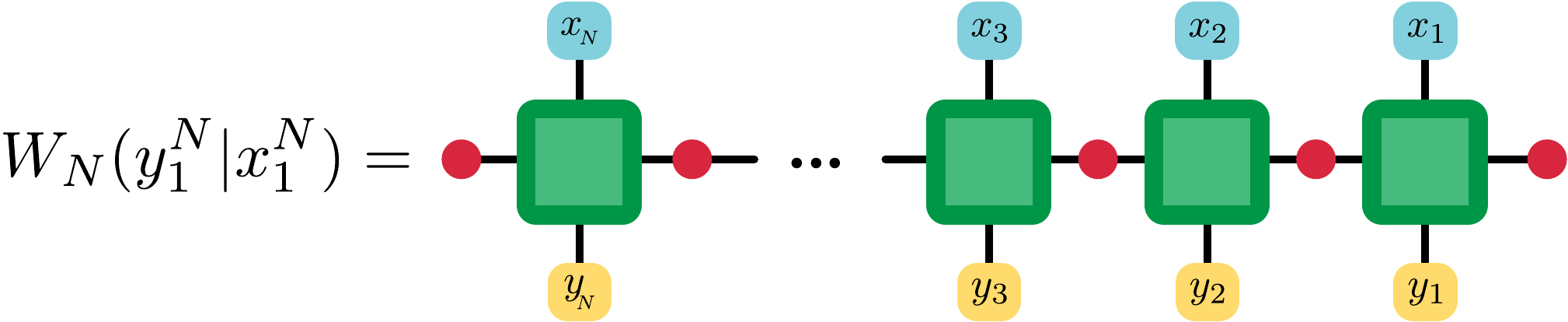}.
\label{eq:WN}
\end{equation}

The rank-2 tensors are the stochastic matrices $q(C'|C)$ defining the channel with memory. For the case of the Gilbert-Elliott channel, this would be
\begin{equation}
\includegraphics[scale=0.5]{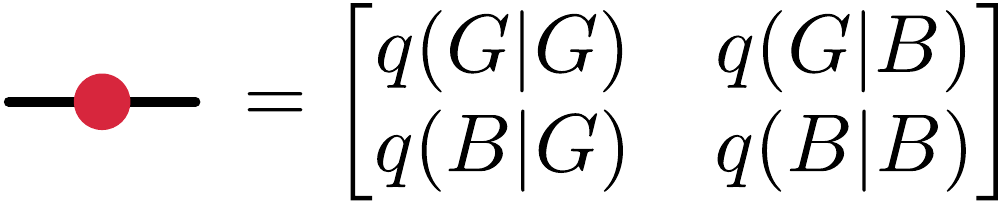}\ .
\label{eq:Wcorr}
\end{equation}

A rank-4 tensor can be thought of as a matrix of matrices, i.e. for fixed values of $i$ and $j$, the object $A_{i,j,k,l}$ is a rank-2 tensor, a.k.a. a matrix. From this perspective, for fixed values $c, c'$ of its horizontal edges, the rank-4 tensors appearing in \eq{WN} is the matrix $\delta_{c,c'} W_c$ where $\delta$ denotes the Kronecker delta and $W_c(y|x) = p(y|x,c)$. In the specific case of the Gilbert-Elliott channel, this would be
\begin{equation}
\includegraphics[scale=0.5]{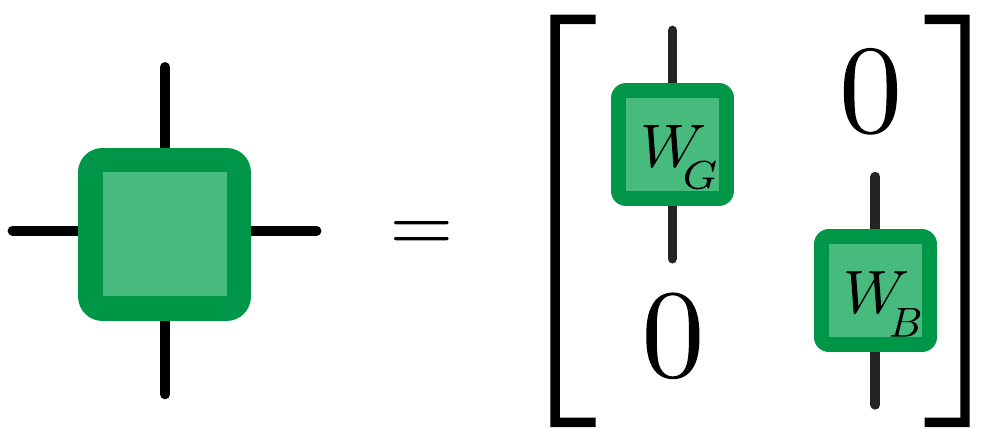}\ .
\label{channel}
\end{equation}

\section*{Acknowledgment}
This work was supported by Canada’s NSERC and the FRQNT. Computations were provided by Compute Canada and Calcul Québec.

\ifCLASSOPTIONcaptionsoff
  \newpage
\fi

\bibliographystyle{ieeetr} 
\bibliography{biblio}

\end{document}